\documentclass[%
 reprint,
nofootinbib,
 amsmath,amssymb,
 aps,
floatfix,
twoside]{revtex4-1}

\usepackage{graphicx}
\usepackage{dcolumn}
\usepackage{bm}
\usepackage[utf8]{inputenc}
\usepackage[portuguese]{babel}
\usepackage[T1]{fontenc}
\usepackage{amsthm}
\usepackage{amsmath}
\usepackage{epigraph}
\usepackage[dvipsnames]{xcolor}

\usepackage{etoolbox}

\makeatletter
\patchcmd{\epigraph}{\@epitext{#1}}{\itshape\@epitext{#1}}{}{}
\makeatother

\begin{document}

\title{
	{\Large A Termodinâmica do Problema do Caixeiro Viajante} \\
	(The Thermodynamics of the Travelling Salesman Problem)}
\author{Paulo J. P. de Souza$^{1}$}

\email[]{paulo.paulino.souza96@gmail.com}

\affiliation{$^1$Departamento de Física, Universidade Federal de São Carlos, São Carlos, SP, Brasil}

\begin{abstract}
Neste trabalho, de intuito pedagógico, revisitamos o formalismo matemático e a interpretação física, baseada na mecânica estatística, da meta-heurística \textit{simulated annealing}. Apresentamos a formulação matemática deste algoritmo de otimização e o porquê dele, de fato, reproduzir a solução ótima ou um boa solução aproximada para o problema em questão. Também, estudamos como o problema do caixeiro viajante se transforma em um problema de cadeias de Markov e, a partir disso, foram feitas simulações do método de \textit{simulated annealing}.
Fizemos as simulações  para o caixeiro viajante em um cenário 50 cidades distribuídas em um círculo e encontramos a solução ótima. 
Além disso, aplicamos o método em uma situação com 100 cidades e uma boa aproximação foi encontrada. 
\begin{description}
\item[Palavras-chave] Cadeias de Markov, Recozimento Simulado,  Caixeiro Viajante.  
\end{description}

\bigbreak 
In this pedagogical work we reviewed the mathematical formalism and the physical interpretation, based on statistical mechanics, of the meta-heuristics called simulated annealing. 
Moreover, we presented the mathematical formulation of the algorithm and why it is capable to yield the optimal solution or a good approximated solution of a given problem. Furthermore, we described the travelling salesman problem, showing its interpretation as a Markov Chain and how the simulated annealing can be used to optimize it and we did its simulations for two scenarios. Firstly, for 50 cities distributed around a circle and we found the best solution. 
Finally, we applied the meta-heuristic in a another instance, with 100 nodes random uniformly distributed in a square, and one shows that it allows finding a good solution.  

\begin{description}
\item[Keywords] Markov Chains, Simulated Annealing,  Traveling Salesman Problem.  
\end{description}
\end{abstract}

\maketitle

\epigraph{"Se as leis físicas fossem pessoas, a Termodinâmica seria a bruxa da vila. Ao longo de três  séculos, ela sorri em silêncio enquanto outras teorias florescem e murcham, sobrevivendo as revoluções da Física. As outras teorias a acham um pouco estranha, de natureza diversa, mas ainda, todas as outras ainda lhe vem pedir conselhos e não ousam contradizê-la".}{-- \textup{Lídia del Rio}}

\section{Introdução}

Problemas de otimização estão relacionados com diversas áreas das ciências exatas, como Física, Engenharia~\cite{engineering}, Biologia~\cite{biology} e Ciência da Computação e da Informação. 
Mais especificamente, a tarefa de encontrar trajetórias de sistemas dinâmicos em mecânica clássica, do ponto de vista do princípio da mínima ação é, essencialmente, um problemas de otimização~\cite{goldstein}. 
Na engenharia, o desenvolvimento dessa classe de problemas, que faz parte da área da pesquisa operacional~\cite{op}, é fundamental na otimização logística de uma indústria, desde sincronização das linhas de produção até o planejamento do transporte.

Em linhas gerais, um problema de otimização se traduz em identificar mínimo ou um máximo, um extremo, de algum objeto matemático que, geralmente, é ou pode ser representado como uma função ou um funcional. 
Os problemas relacionados à extremização podem ser divididos em duas categorias, problemas contínuos e problemas discretos. 
Um exemplo de problema contínuo é encontrar os vértices de uma função.
Já em problemas discretos, um exemplo é encontrar o menor caminho entre dois nós de um grafo.
Dentro da categoria de problemas de otimização discretos existem os problemas NP, tempo polinomial não determinístico, nos quais a busca pela solução ótima é feita em um espaço de soluções que cresce de forma não polinomial - em geral de forma exponencial ou fatorial - com o tamanho da entrada problema. 
Desse modo, a resolução exata desses problemas necessita de uma grande quantidade de recursos computacionais, o que traz a necessidade de se encontrar maneiras de buscar boas soluções aproximadas. 
Foram desenvolvidas diversas estratégias e heurísticas para abordar de maneira aproximada essa família de problemas.
Por exemplo, temos os algoritmos genéticos~\cite{holland1992adaptation}, o algoritmo guloso~\cite{goldbarg} e métodos de \textit{cross-entropy}~\cite{rubinstein1999cross}.
Neste trabalho, analisaremos os problemas cujas formulações matemáticas podem ser escritas como uma de cadeia de Markov~\cite{tome2014dinamica, markov}\footnote{Cadeias de Markov são processos estocásticos ditos sem memória, isso é, sua evolução para o próximo estado só depende do estado atual.} com o foco voltado para o entendimento do método \textit{simulated annealing}, SA, em português: recozimento simulado~\cite{kirkpatrick1983optimization}. 

O SA, é uma meta-heurística\footnote{Método heurístico para resolver problemas de otimização de forma genérica} comumente usada em problemas NP derivados de problemas de otimização combinatória,  e/ou quando o custo benefício da solução aproximada é preferível em relação à solução ótima~\cite{2019simulated}. 
Essa meta-heurística é um método estocástico para aproximação de um mínimo global de uma função discreta chamada de função objetiva. 
Ele consiste em transformar o problema de otimização em um problema termodinâmico definindo as regras do ambiente, como os microestados, suas respectivas energias e a regra de transição entre eles.
O sistema termodinâmico é iniciado com uma temperatura finita e o papel do algoritmo é diminuir a temperatura e termalizar o sistema repetidas vezes até que temperaturas próximas do zero absoluto sejam atingidas. 
Em teoria, a solução ótima do problema de otimização é encontrada quando a temperatura do sistema termodinâmico chega ao zero absoluto e o sistema ao seu estado fundamental. 
Além de ser um bom método de otimização~\cite{vcerny1985thermodynamical}, o estudo do SA se justifica pela sua rica interpretação termodinâmica que pode ser usada para melhor compreender sistemas termodinâmicos estocásticos e, também, pelo fato que ele compõe outros algoritmos heurísticos mais sofisticados e mais eficientes~\cite{henderson2003theory}. 

Um problema de otimização NP que pode ser abordado via SA é o problema do caixeiro viajante, do inglês, \textit{travelling salesman problem} (TSP)~\cite{traveler, TSP} que consiste em encontrar a melhor rota que passa em todas as cidades de uma lista apenas uma vez e retorna à cidade inicial.
Sua importância se dá pelas suas diversas aplicações como o gerenciamento de chegadas e partidas em aeroportos, sequenciamento de DNA ou logística de transporte~\cite{matai}. 

O estado da arte dos métodos exatos para o TSP foi, utilizando um algoritmo de programação dinâmica com complexidade $\mathrm{O}(n^2 2^N)$~\cite{held1962dynamic}, a obtenção da solução ótima para 85,900 cidades, o que levou 139 CPU-anos\footnote{Um CPU segundo é o tempo para realizar 1 giga de operações de ponto flutuante} para ser encontrada em um supercomputador~\cite{2009certification}. 
No contexto da computação quântica, há um algoritmo que possui complexidade, para casos onde o número de conexões das cidades é limitado, até quadraticamente menor que aquela do algoritmo clássico~\cite{moylett2017quantum}.
Alguns outros algoritmos exatos podem ser encontrados na referência~\cite{woeginger2003exact}.
Já os melhores algoritmos heurísticos encontram soluções próximas da solução ótima com erro próximo a  $5\%$ para problemas com 3 milhões de cidades em poucas horas de tempo de cpu~\cite{heuristics}. 

O objetivo desse trabalho é justificar matematicamente e entender a interpretação física, pela ótica da mecânica estatística, da meta-heurística SA e, além disso, compreender seu comportamento e sua execução. Para isso, começamos este trabalho definindo um processo estocástico e quando ele se torna um processo markoviano.
Depois, demonstramos que todo processo markoviano caminha para a maximização da entropia em um intervalo finito de tempo e, dessa forma, para uma distribuição estacionária de estados, o que significa atingir o equilíbrio termodinâmico. 
Após isso, usamos a teoria estatística desenvolvida para mostrar que  a diminuição da temperatura de forma lenta o suficiente (adiabática) leva o sistema ao estado fundamental.
Na segunda parte formulamos o problema do caixeiro viajante como um problema termodinâmico e, também, discutimos como o problema do caixeiro viajante, nessa formulação, é visto como um problema de Ising~\cite{libero2000ising, baxter}, portanto, tudo o que é aqui discutido se estende, em geral, aos problemas tipo-Ising\footnote{Problemas tipo-Ising são aqueles que podem ser escritos como uma função da combinação variáveis binárias.}.
Por fim, implementamos a simulação do SA aplicado ao TSP e apresentamos uma discussão dos resultados.

Fizemos as simulações usando a linguagem python~\cite{python}, a biblioteca numérica numpy~\cite{numpy} e para as figuras a biblioteca matplotlib~\cite{matplotlib} e o software Inkscape~\cite{inkscape}. 
O código das simulação pode ser encontrado no repositório público~\cite{git}.

\section{Métodos}

\subsection{Cadeias de Markov} 
Um processo estocástico é definido por uma família ou conjunto de variáveis aleatórias $\left\{x_t,\, t\in T=\mathbb{R}^*\right\}$ que evoluem de forma ordenada, o que traz a possibilidade de visualizá-las como um ordenamento temporal de um sistema probabilístico. 
Podemos tomar de exemplo de processos estocásticos a série temporal da cotação do dólar, turbulência ou ruídos térmicos ou eletromagnéticos~\cite{stochasticProcess}.
As variáveis aleatórias podem ser o valor atual do dólar, um microestado de uma sistema de mecânica estatística ou valor da face de um dado em uma sequência de lançamentos. 

Neste trabalho vamos considerar apenas processos estocásticos onde tanto o tempo como a variável aleatória podem ser discretizados. 
Para essa situação, o processo estocástico é totalmente determinado pela distribuição de probabilidade conjunta das variáveis aleatórias assumirem os valores $\left\{x(t_i)=n_i,\; i \in \left\{1, 2, \cdots, l\right\} \right\}$, 
\begin{equation}
	p_l\left(n_1, n_2,\cdots, n_l \right),
\end{equation}
onde $l$ é o número de intervalos temporais. 
A distribuição condicional
\begin{equation}
	p_{l}\left(n_{l}|n_1, n_2, \cdots, n_{l-1} \right),
\end{equation}
fornece a probabilidade da variável aleatória assumir o valor $n_{l}$ dado que ela assumiu a sequência de eventos $n_0, n_1, \cdots, n_{l-1}$.
Se o processo estocástico possui a propriedade 
\begin{equation}\label{eq:markovian}
	p_{l}\left(n_{l}|n_1, n_2, \cdots, n_{l-1} \right) = p_{l}\left(n_{l}|n_{l-1}\right)
\end{equation}
dizemos que ele é markoviano. 
Podemos descrever essa propriedade como uma ausência de memória, isso é, o sistema não leva em conta os eventos anteriores, de modo que o passo seguinte só depende do valor atual da variável aleatória. 
Usando a equação~(\ref{eq:markovian}) podemos expressar $p_l(n_l)$ independentemente dos valores assumidos anteriormente 
\begin{equation}
	p_l(n_l) = \sum\limits_{n_{l-1}} p_{l}\left(n_{l}|n_{l-1}\right)p_{l-1}\left(n_{l-1}\right).
\end{equation}
A probabilidade condicional $p_l(n_{l}|n_{l-1})$ é interpretada como a probabilidade de transição entre os estados $n_l$ e $n_{l-1}$. 
Em geral essa probabilidade de transição é dependente do tempo, mas se não o for, temos um processo markoviano independente do tempo e assim podemos escrever
\begin{equation}
	p_{l}(n_{l}|n_{l-1}) = T(n_{l}, n_{l-1}), 
\end{equation}
onde $T(n_{l}, n_{l-1})$ é a probabilidade de transição entre os estados $n_{l}$ e $n_{l-1}$.
Dessa forma, renomeando as variáveis $n \equiv n_l$ e $m \equiv n_{l-1}$, a probabilidade da variável aleatória assumir o valor $n$ no intervalo de tempo $t_i=l$ é dada por
\begin{equation}\label{eq:markoviano2}
	p_{l}(n) = \sum\limits_{m}T(n, m)p_{l-1}(m).
\end{equation}

É possível interpretar $T(n,m)$ como um elemento de uma matriz $\hat{T}$ que é chamada de \textit{matriz de transição} ou \textit{matriz estocástica} que deve obedece às seguintes propriedades: 
\begin{enumerate}
	\item $T(n,m) \geq 0,\; \forall\; m,n$;
	\item $\sum\limits_n T(n,m) = 1$.
\end{enumerate}
Essas propriedades são as condições axiomáticas de probabilidade, que são a não existência de probabilidades negativas e a condição de normalização. 
Notamos que as colunas, segundo índice,  da matriz de transição representam os estados instantâneos e as linhas, primeiro índice, os possíveis estados seguintes, de forma que $T(n,m)$ representa a probabilidade do sistema que está no estado $m$ ir para o estado $n$. 
Se definirmos uma matriz coluna 
\begin{equation}
	\left| P_l \right> = 	\begin{bmatrix}
				p_l(1) \\
  				p_l(2) \\
  				\vdots \\
  				p_l(z) 
			\end{bmatrix},	
\end{equation} 
que possui a probabilidade de se encontrar o sistema em cada estado no instante $t=l$, então a equação~(\ref{eq:markoviano2}) fica escrita como
\begin{equation}
	\left|P_l\right> = \hat{T} \left|P_{l-1}\right>. 
\end{equation}
Das propriedades de $\hat{T}$ e da markovianidade do processo estocástico, pelo teorema de Perron-Frobenius~\cite{perronFrobenius} temos que 
\begin{equation}\label{eq:stationary}
	\left|P\right>= \hat{T} \left|P\right>,
\end{equation}
onde $\left|P\right>$ é matriz de densidade clássica estacionária do sistema. 
A probabilidade estacionária de um estado satisfaz à equação 
\begin{equation}
	\begin{split}
		p(n) = \sum\limits_m T(n,m)p(m) \rightarrow \\
		\rightarrow \sum\limits_m \left\{T(n,m)p(m) - T(m,n)p(n)\right\}=0,
	\end{split}
\end{equation}
porque $\sum_m T(m,n)=1$. 
Se temos uma situação onde cada parcela é igual a zero, então
\begin{equation}\label{eq:detailedBalance}
 	T(n,m)p(m) - T(m,n)p(n)=0,
\end{equation}
e dizemos que a probabilidade estacionária satisfaz a condição de balanço detalhado ou de reversibilidade microscópica~\cite{onsager1931reciprocal, tome2014dinamica}.

\subsection{Entropia}

Na seção anterior descrevemos um processo estocástico arbitrário, agora vamos fazer uma conexão desses processos com a teoria de informação e com a Termodinâmica. 
Para tal fim, em primeiro lugar, vamos considerar uma função convexa arbitrária, $f(x)$, dotada da propriedade
\begin{equation}\label{eq:convex}
	f(\sum\limits_{m=0} x_m p_m) \leq \sum_{m=0}^N p_m f_m(x_m),
\end{equation}
onde $p_m$ obedece a uma distribuição de probabilidade. 

Em paralelo, podemos escrever, usando a equação~(\ref{eq:markoviano2}) e a equação~(\ref{eq:detailedBalance}), 
\begin{equation}\label{eq:DB2entropy}
	\begin{split}
		p_{l+1}(n) =\sum\limits_{m}T(m,n)\frac{p(n)}{p(m)}p_{l}(m) ,\\
		\frac{p_{l+1}(n)}{p(n)} = \sum\limits_{m}T(m,n)\frac{p_l(m)}{p(m)}. 
	\end{split}
\end{equation}
Agora, substituindo a equação~(\ref{eq:DB2entropy}) na equação~(\ref{eq:convex}) com $p_m = T(m,n)$ e $x_m = p_l(m)/p(m)$ obtemos 
\begin{equation}
	f\left(\frac{p_{l+1}(n)}{p(n)}\right) \leq \sum\limits_m T(m,n) f\left(\frac{p_{l}(m)}{p(m)}\right).
\end{equation}
Por fim, multiplicando por $p(n)$ e somando em $n$
\begin{equation}
	\begin{split}
	\sum\limits_n p(n) f\left(\frac{p_{l+1}(n)}{p(n)}\right) \leq \\
	\leq \sum\limits_n \sum\limits_m p(n) T(m,n) f\left(\frac{p_{l}(m)}{p(m)}\right) = \\
	=  \sum\limits_m  f\left(\frac{p_{l}(m)}{p(m)}\right)  \sum\limits_n p(n) T(m,n) = \\
	= \sum\limits_m p(m) f\left(\frac{p_{l}(m)}{p(m)}\right) . 
\end{split}
\end{equation}

Com esse resultado em mãos podemos prosseguir para a Termodinâmica. 
A entropia de Shannon é definida por 
\begin{equation}\label{eq:shannon}
	S = -k_B\sum_i p_i \ln p_i,
\end{equation} 
onde $k_B$ é a constante de Boltzmann. 
Dessa forma, se substituirmos $f(x) = x \ln x$, então 
\begin{equation}\label{eq:entropyIncrease}
	\begin{split}
	\sum\limits_n p_{l+1}(n) \ln\frac{p_{l+1}(n)}{p(n)} \leq  \sum\limits_m p_l(m) \ln\frac{p_{l}(m)}{p(m)} \\
	\therefore\quad   S_{l+1} \geq S_l,
\end{split}
\end{equation}
e obtemos a conexão desejada entre o processo estocástico markoviano e a Termodinâmica. 
Em complemento, podemos notar que quando sistema atinge a condição estacionária, a desigualdade da equação~(\ref{eq:entropyIncrease}) se torna uma igualdade pois os logaritmos se reduzem a zero. 

Agora, podemos dizer que um processo markoviano, que respeita a condição de balanço detalhado, aumenta a entropia do sistema até o equilíbrio térmico, como rege a segunda lei da Termodinâmica. 
Apesar deste procedimento nos dizer que o sistema caminha para uma configuração de maior entropia, ele não nos fornece o quão próximo estamos desse máximo global.
Tal característica é presente em algoritmos heurísticos e é uma das causas dos seus problemas relacionados à convergência para a solução ótima. 

Em suma, a formulação estocástica aqui descrita possui um profundo sentido termodinâmico e, com isso, podemos estudá-la através dessa ótica. 

\subsection{Monte Carlo}
Os métodos de Monte Carlo, MC, são um conjunto de técnicas computacionais que utilizam de amostragens probabilísticas para resolver problemas numéricos~\cite{monte}. Os algoritmos de \textit{annealing} fazem parte destas técnicas, pois se baseiam na amostragem das transições dos estados do sistema.

Vimos a descrição de cadeias de Markvov como um objeto matemático abstrato, agora vamos ver sua aplicação em problemas termodinâmicos. 
Para construir um algoritmo de MC temos que partir de uma regra para amostragem, ou seja, devemos ter uma forma de gerar aleatoriamente estados do sistema.
No nosso contexto, devemos ter uma regra para transitar entre estados do sistema, isso é, obter uma matriz estocástica. 
Para tal objetivo, vamos considerar a função de partição canônica, que nos diz que a probabilidade de encontrar o sistema no microestado $s$, no equilíbrio termodinâmico, é dada pela expressão 
\begin{equation}\label{eq:prob}
	p(s) = \frac{1}{Z}e^{-\beta H(s)},
\end{equation}
onde $H(s)$ é a Hamiltoniana (função energia) do sistema para a configuração $s$, $Z=\sum_s e^{-\beta H(s)}$ é a função de partição e $\beta=1/k_B T$, sendo $T$ a temperatura. 

Continuando, o nosso objetivo, para o MC,  é encontrar uma matriz de transição $T$ e atingi-lo-emos resolvendo a equação 
\begin{equation}\label{eq:monteCarloDB}
	\sum\limits_{s'}T(s,s')p(s') = p(s). 
\end{equation}
Substituindo a equação~(\ref{eq:prob}) na equação~(\ref{eq:monteCarloDB}) ficamos com 
\begin{equation}
	T(s, s') = e^{-\beta\left(H(s)-H(s')\right)}, 
\end{equation}
mas se $H(s') > H(s)$ teríamos um problema, pois a probabilidade seria maior que um. 
Usando uma abordagem \textit{ad hock} podemos redefinir $\hat{T}$ como 
\begin{equation}\label{eq:transitionMatrix}
  T(m, n) = 
  \begin{cases}
	  	\frac{1}{N}e^{-\beta\left[H(m)-H(n)\right]},\; \text{para } H(m) > H(n) \\
		\frac{1}{N},\; \textnormal{caso contrário} 
  \end{cases},
\end{equation}
onde $N$ é o número de estados acessíveis do microestado $m$.
Notamos que $\hat{T}$ é função de $\beta$ e que obedece às propriedades de balanço detalhado, normalização e positividade\footnote{A matriz de transição encontrada não é única, pois a equação~(\ref{eq:monteCarloDB}) admite outras soluções.}.
Além disso, da equação~(\ref{eq:transitionMatrix}) extraímos que todos os estados do sistema são atingíveis em um intervalo finito de tempo, pois todos os elementos da matriz estocástica são diferentes de zero, já que a energia é definida e finita para todos os estados. 

Apesar da equação~(\ref{eq:prob}) descrever apenas sistemas em equilíbrio, podemos usá-la para construir as transições entre estados, mesmo fora do equilíbrio, pois a matriz $\hat{T}$ encontrada obedece às propriedades de matriz estocástica e isso é suficiente para que ela leve o sistema ao equilíbrio termodinâmico. 
O parâmetro $\beta$ possui unidade do inverso de energia, dessa forma, ele vai impor uma escala de energia para as flutuações que equivalem às  probabilidades de transição entre microestados do sistema.
Em uma primeira análise, temos que se $\beta$ é muito grande, as flutuações são muito baixas e não há transições entre os estados. Por outro lado, se $\beta$ é muito pequeno todos os estados são equiprováveis.

\begin{figure}
	\centering 
		\includegraphics[scale=0.28]{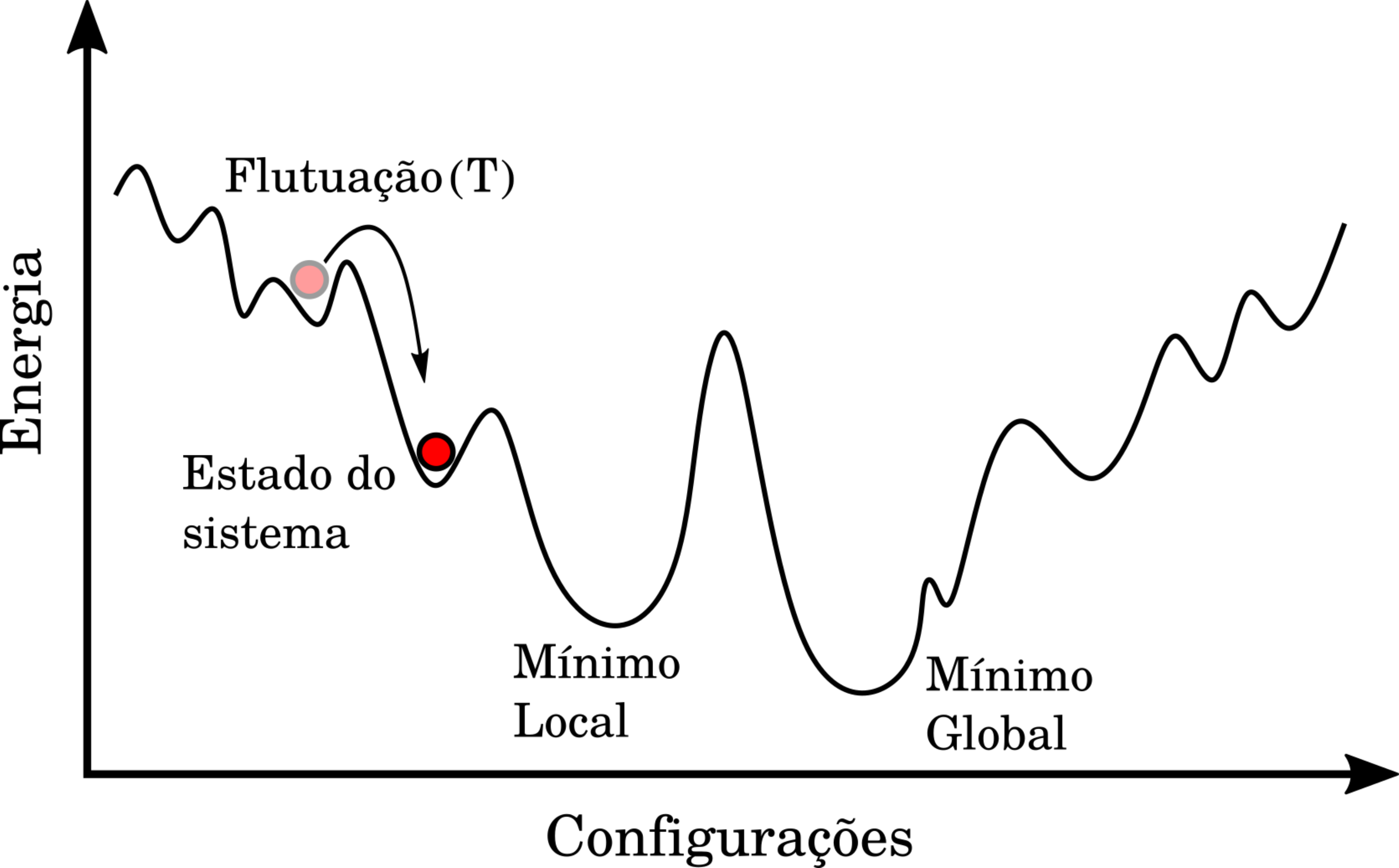}
		\caption{Uma representação esquemática do método \textit{simulated annealing}. O eixo $x$ representa as configurações e o eixo $y$ suas respectivas energias. O círculo vermelho é identificado como a configuração do sistema em um dado instante e o círculo opaco é a configuração imediatamente anterior. A flecha ligando os dois círculos representa a flutuação termodinâmica que levou o sistema de um estado ao outro ``escalando'' os picos da superfície.  Notamos que se a energia da flutuação não for grande o suficiente o sistema fica preso em mínimos locais e se for muito grande ele não se estabiliza em nenhum vale.}
		\label{fig:energias&config}
\end{figure}

A figura~(\ref{fig:energias&config}) mostra um espaço de estados arbitrário, onde temos a energia das configurações no eixo vertical e no eixo horizontal as configurações vizinhas, que são atingíveis em um intervalo de tempo.  
Podemos observar que o sistema caminha, escala e desce os vales da superfície de energia dos microestados pelo efeito das flutuações térmicas. 
\subsection{O estado fundamental}

Para entender como o algoritmo reproduz o estado fundamental do sistema termodinâmico vamos considerar alguns fatos da mecânica estatística e da termodinâmica. 
Mostramos que ao inicializarmos o nosso algoritmo MC com uma temperatura não nula e realizarmos a dinâmica da cadeia de Markov, em um número finito de passos o sistema caminhará para a situação de maior entropia daquela temperatura, i.e. atingirá o equilíbrio termodinâmico. 
Se com o sistema no equilíbrio termodinâmico abaixarmos a sua temperatura por um $\delta T$ pequeno, mas ainda não sendo um infinitésimo, o sistema sairá do equilíbrio, mas com o decorrer da dinâmica retornará a ele. 
Podemos realizar essa operação repetidamente até chegarmos em uma temperatura tão próxima quanto quisermos do zero absoluto e já nesse cenário, podemos recorrer à terceira lei da termodinâmica. A terceira lei nos diz que a entropia no zero absoluto é mínima e o sistema está no seu estado fundamental.
Dessa forma, podemos concluir que, idealmente, o sistema levado a temperaturas próximas de zero de forma suficientemente lenta, também será levado ao estado fundamental, ou seja, até a solução ótima do problema de otimização.

\subsection{O problema do Caixeiro Viajante}

O problema do caixeiro viajante é anunciado pela pergunta: \textit{Dada uma lista de cidades, qual a rota que otimiza o custo de uma viagem que passa em todas as cidades apenas uma vez e retorna à cidade inicial?}

Começaremos fazendo a tradução da linguagem usada para tratar processos estocásticos para o problema do caixeiro viajante. 
Anteriormente, tínhamos falado sobre uma variável aleatória que em cada intervalo temporal pode assumir um valor aleatório.
\begin{figure*}
	\begin{center}
	\includegraphics[scale=0.4]{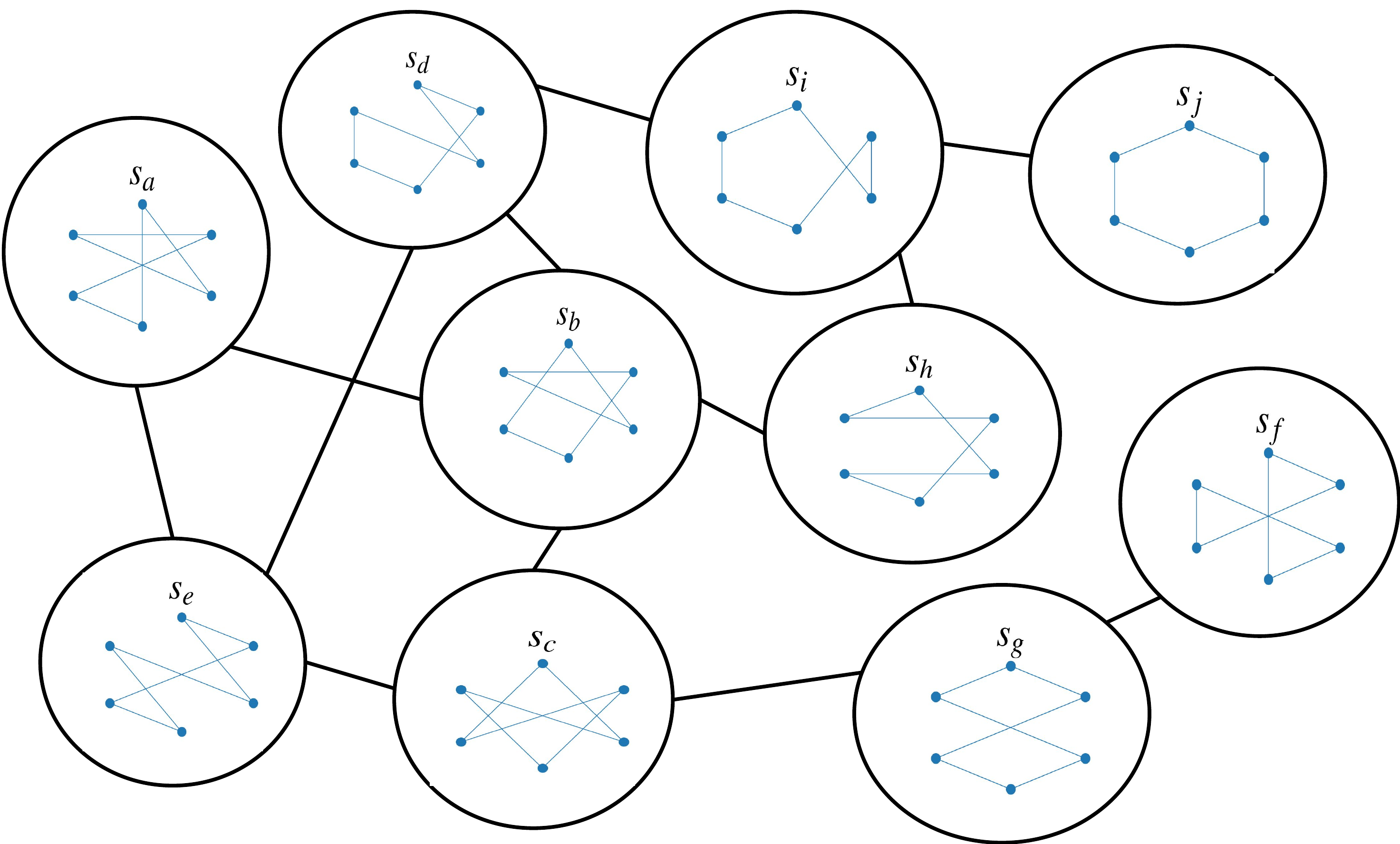}
	\caption{Grafo representando os estados do processo estocástico para uma caso com 6 cidades dispostas em um hexágono, isso é, as trajetórias e suas vizinhanças. Cada vértice representa uma lista de cidades e as arestas conectam trajetórias que distam de uma permutação na ordem das visitas. Dessa forma, se a configuração de início for a $S_a$, o sistema pode se mover para a configuração $S_e$ ou para a $S_b$, com isso, observamos que partindo de $S_a$, um caminho possível até o estado fundamental é $\left\{S_a, S_b, S_h, S_i, S_j \right\}$. Note que há uma independência da cidade inicial, do sentido em que as cidades são visitadas e que, também, há simetria na disposição de cidades, e por isso foram excluídos os estados degenerados, pois estes geram uma rede análoga a essa.}
	\label{fig:statesNetworks}
	\end{center}
\end{figure*}
A figura~(\ref{fig:statesNetworks}) representa todos os possíveis estados do sistema, em outras palavras, todas as possíveis rotas ou trajetórias. 
Suponhamos que no instante inicial o sistema esteja na configuração correspondente à rota $S_a$ e que o nosso processo estocástico faça apenas uma permutação no vetor de cidades a cada passo (o que é equivalente a cada instante de tempo). 

Desse modo, a cada passo de execução do algoritmo a configuração do sistema pode ser alterada somente para os estados $S_e$ e $S_b$ com as probabilidades dadas pela matriz de transição, equação~(\ref{eq:transitionMatrix}). 
Se quisermos saber a probabilidade de encontrar o sistema no estado $S_j$ basta usar a equação~(\ref{eq:markoviano2}).
Uma forma de pensar neste problema é fazendo uma analogia com o caminhante aleatório que aqui se movimenta pelo grafo que representa as rotas, de acordo as probabilidades dadas pela matriz estocástica. 

\subsection{Formulação Hamiltoniana}

Podemos fazer a transcrição termodinâmica do problema do caixeiro viajante definindo a chamada função objetiva ou Hamiltoniana. 
Antes, formalizaremos os aspectos do problema definindo o espaço de configurações como 
\begin{equation}
	\mathcal{S} = \left\{s_i\right\}_\Omega, \quad s_i = \left(c_0, c_1, \cdots, c_{l-1}\right),
\end{equation}
onde $s_i$ é um estado do sistema, uma rota, e $c_i$ é a $i$-ésima cidade visitada. 
A energia do sistema é definida como a função
\begin{equation} 
	H: \mathcal{S} \longmapsto \mathbb{R}, 
\end{equation}
com 
\begin{equation}
	H(s) = D(c_0, c_{N-1}) + \sum\limits_{i=0}^{N-1} D(c_i, c_{i+1}),
\end{equation}
onde $D(c_i, c_j)$ é a distância, ou custo, entre a cidade $i$ e a cidade $j$.
Para o TSP, como condição de contorno, temos que não se pode visitar a mesma cidade duas vezes, que todas as cidades devem ser visitadas e que o caminhante deve retornar à cidade de início. 
Essas condições nos dizem que para uma dada trajetória $ s = \left[c_0, c_1, \cdots, c_{N-1}\right]$ a força randômica, temperatura, faz permutações do tipo $c_j \rightarrow c_i$ e $c_i \rightarrow c_j$ com $i\neq j$.
Estamos diante de um processo markoviano porque as probabilidades não variam com o tempo, todo estado $s\in \mathcal{S}$ pode ser atingido em um espaço de tempo finito, os microestados possíveis no passo seguinte só dependem do atual e a condição de balanço detalhado, equação~(\ref{eq:detailedBalance}), é respeitada. 
Agora podemos descrever o algoritmo \textit{simulated annealing}, que reunirá todos os atributos anteriormente descritos, através dos seguintes passos: 
\begin{enumerate}
	\item Defina um estado inicial aleatório; 
	\item Faça uma permutação aleatória na lista de cidades e defina o novo estado $s'$;
	\item Se $H(s') < H(s)$: aceite o novo estado;
	\item Senão: sorteie um número aleatório $\xi$, entre $0$ e $1$. Se $\xi < T(s', s)$ aceite o estado $s'$, senão descarte $s'$.
	\item A cada $\eta$ passos temporais dados reduza a temperatura, $T_{i+1} = \alpha T{i},\quad \alpha \in \left\{0<x<1,\; x \in \mathrm{R}\right\}$\footnote{Aqui foi dado um exemplo de resfriamento, mas existem outras formas};
	\item Mantenha o estado de menor energia salvo;
	\item Estabeleça uma condição de parada, como o número de passos dados. 
\end{enumerate}
Esses passos são muito similares ao algoritmo Metropolis~\cite{hastings1970monte}, porém o item $5$, a diminuição da temperatura, o identifica como característica do \textit{simulated annealing}. 
A ideia do item 5 é que a temperatura seja reduzida pouco a pouco e somente após o sistema estar termalizado, isso é, ter sua função entropia em um máximo. 
Isso, em teoria, é necessário para que haja uma redução térmica próxima à adiabática e, assim, evitamos reduzir a temperatura com o sistema em estados metaestáveis (mínimos locais). 

Com isso, podemos resumir a relação do problema do caixeiro viajante com a termodinâmica pelos seguintes itens:
\begin{enumerate}
	\item Cidades e conexões $\rightarrow$ Sistema termodinâmico;
	\item Trajetória do caixeiro viajante $\rightarrow$ microestado do sistema ou configuração do sistema;
	\item Distância da trajetória $\rightarrow$ energia da configuração;
	\item Otimização da viagem $\rightarrow$ sistema no estado fundamental (temperatura próxima de zero).
\end{enumerate}

\subsection{O Caixeiro Viajante como um problema de Ising}

O modelo de Ising é um modelo para sistemas de spins comutantes, muito importante em matéria condensada e mecânica estatística, para o entendimento da teoria e explicação de alguns fenômenos como ferro e paramagnetismo e classes de universalidades. 
No contexto desse artigo há a referência~\cite{libero2000ising} e em um contexto mais geral e aprofundado pode-se consultar a referência~\cite{baxter}.
Com a intenção de melhorar o nosso entendimento do TSP como um problema termodinâmico vamos olhar como ele pode ser mapeado em um problema tipo-Ising. 
Definindo a variável binária, spin, como  
\begin{equation}
	n_{i, \alpha} \in \left\{0,1 \right\}, 
\end{equation}
que assume $1$ quando o caixeiro passa pela $i$-ésima cidade no $\alpha$-ésimo passo e $n_{i, \alpha}=0$ caso contrário. 
Para um sistema com $l$ cidades, nossas condições de contorno são
\begin{equation}
	\sum\limits_{i}^l n_{i,\alpha} = \sum\limits_{\alpha}^l n_{i,\alpha} = 1, \quad \forall i, \alpha
\end{equation}
isso, porque ele não pode passar na mesma cidade duas vezes e ele só pode estar em uma cidade por um dado intervalo de tempo. 
O acoplamento entre dois spins é dado por 
\begin{equation}
	J_{i, j} = D(i,j),
\end{equation}
onde $D(i,j)$ é a distância entre as cidades $i$ e $j$. 
Finalmente, o comprimento da viagem $s$ é descrito como
\begin{equation}
	\begin{split}
	H(s) = \sum\limits_{\alpha=1}^l \sum\limits_{1<i,j<l} J_{i,j}n_{i,\alpha} n_{j, \alpha+1}=\\
	= \frac{1}{4}\sum\limits_{\alpha=1}^l \sum\limits_{1<i,j<l} J_{i,j} \sigma^z_{i,\alpha}\sigma^z_{j, \alpha+1} + \sum\limits_{1<i,j<l} J_{i,j} \sigma^z_{i, \alpha} + \text{const},
\end{split}
\end{equation}
sendo $\sigma_{i, \alpha}^z=2n_{i, \alpha}-1$ a variável Ising. 
Essa equação é a Hamiltoniana para um modelo de Ising com interações aleatórias e com campo magnético não-hemogêneo.
Note que para fazer o mapeamento de $N$ cidades é necessário $N^2$ variáveis de spin, pois precisamos de uma lista de cidades para cada um dos $N$ intervalos de tempo.
Como essa Hamiltoniana tipo-Ising possui as mesmas características matemáticas da Hamiltoniana do caixeiro viajante, podemos usar toda sua construção aqui feita para abordar qualquer problema de otimização que recaia em um problema tipo-Ising~\cite{lucas2014ising}. 

\section{Resultados}

Com o intuito de ilustrar como é o comportamento e funcionamento do processo de \textit{annealing}, o simulamos para o caixeiro viajante. Consideramos nas simulações o problema do caixeiro viajante completo, todas as cidades estão conectadas entre si, e  simétrico, a distância da cidade $i$ para a cidade $j$ é a mesma da cidade $j$ para a cidade $i$. 
Ademais, como discutido na referência~\cite{henderson2003theory} a topologia das cidades influencia na convergência, pois em casos onde há mais mínimos locais ela se da mais lentamente.
Dito isso, consideramos duas topologias, uma com as cidades colocadas circularmente (menos mínimos locais) e outra onde as cidades são distribuídas aleatoriamente em um quadrado (mais mínimos locais).

\begin{figure}
	\begin{center}
	\includegraphics[scale=0.50]{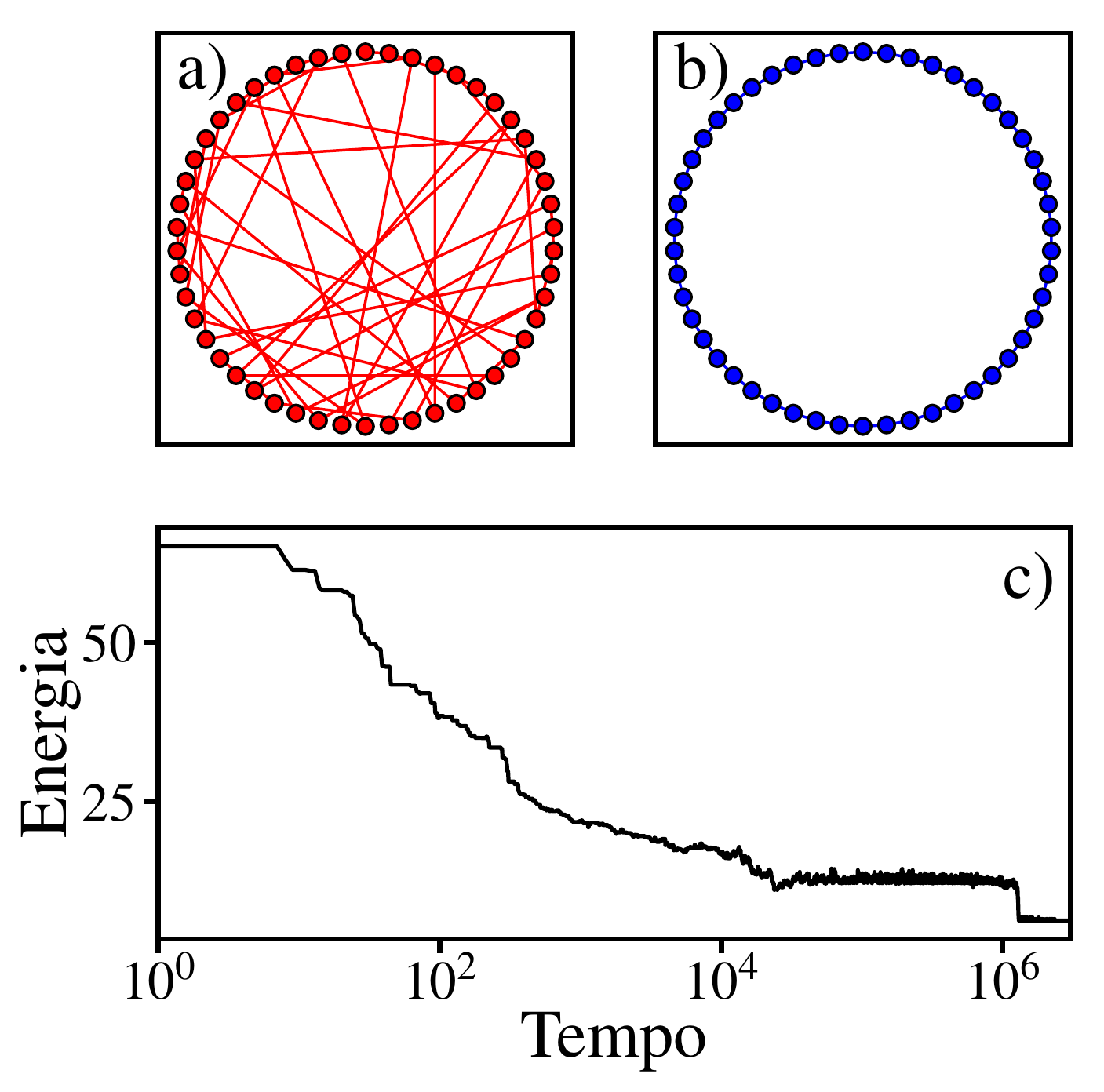}
	\caption{Simulação do problema do caixeiro viajante com 50 cidades distribuída em um circulo de raio 1 com $T_{\text{inicial}}=10$, $\alpha=0,995$ e com o abaixamento da temperatura feito a cada $10^4$ passos. A figura a) corresponde ao estado inicial, $\textnormal{distância}=65,14$,  e a figura b) representa o estado final, distância $=6,27$, da dinâmica com $3\times10^6$ passos, onde é encontrado o estado fundamental. A figura c) mostra a evolução da energia do sistema ao longo da execução do algoritmo \textit{simulated annealing}.}
	\label{fig:circ}
	\end{center}
\end{figure}

A figura~(\ref{fig:circ}) mostra um caso que 50 cidades estão distribuídas em um circulo e onde conhecemos a solução ótima, que é a configuração circular. 
A figura~\ref{fig:circ}(b), mostra que o algoritmo  encontrou a solução ótima com aproximadamente $3\times10^6$ iterações. 
Um outro ponto notável está na figura~\ref{fig:circ}(c), analisando-a vemos que próximo ao Tempo = $10^6$ há um descida brusca na energia o que decorre do salto de um estado metaestável para um estado de menor energia. 

\begin{figure}
	\begin{center}
	\includegraphics[scale=0.38]{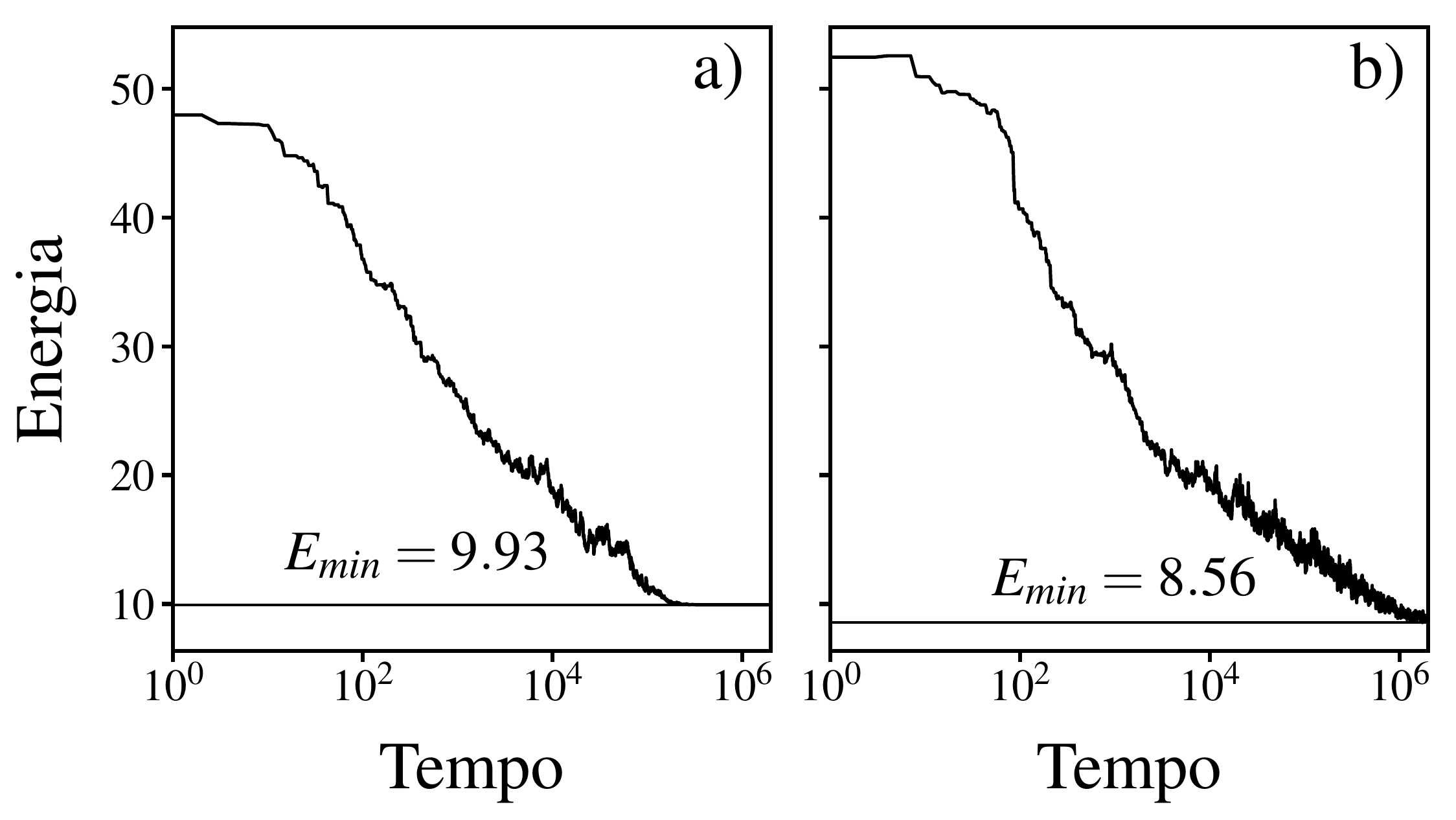}
	\caption{Energia do sistema em função do número de iterações ($\tau$) do algoritmo \textit{simulated annealing} para um sistema com 100 cidades distribuídas pela densidade uniforme de probabilidade. Em a) foi usado $\beta_{\text{inicial}}=10$,  $\alpha=0,90$ e a temperatura foi decrescida a cada $10^4$ passos. Em b) foi usado $\beta_{\text{inicial}}=10$,  $\alpha=0,99$ e a temperatura foi decrescida a cada $10^4$ passos. Note que o eixo $x$ está em escala logarítmica.}
	\label{fig:eList}
	\end{center}
\end{figure}

\begin{figure}
	\begin{center}
	\includegraphics[scale=0.42]{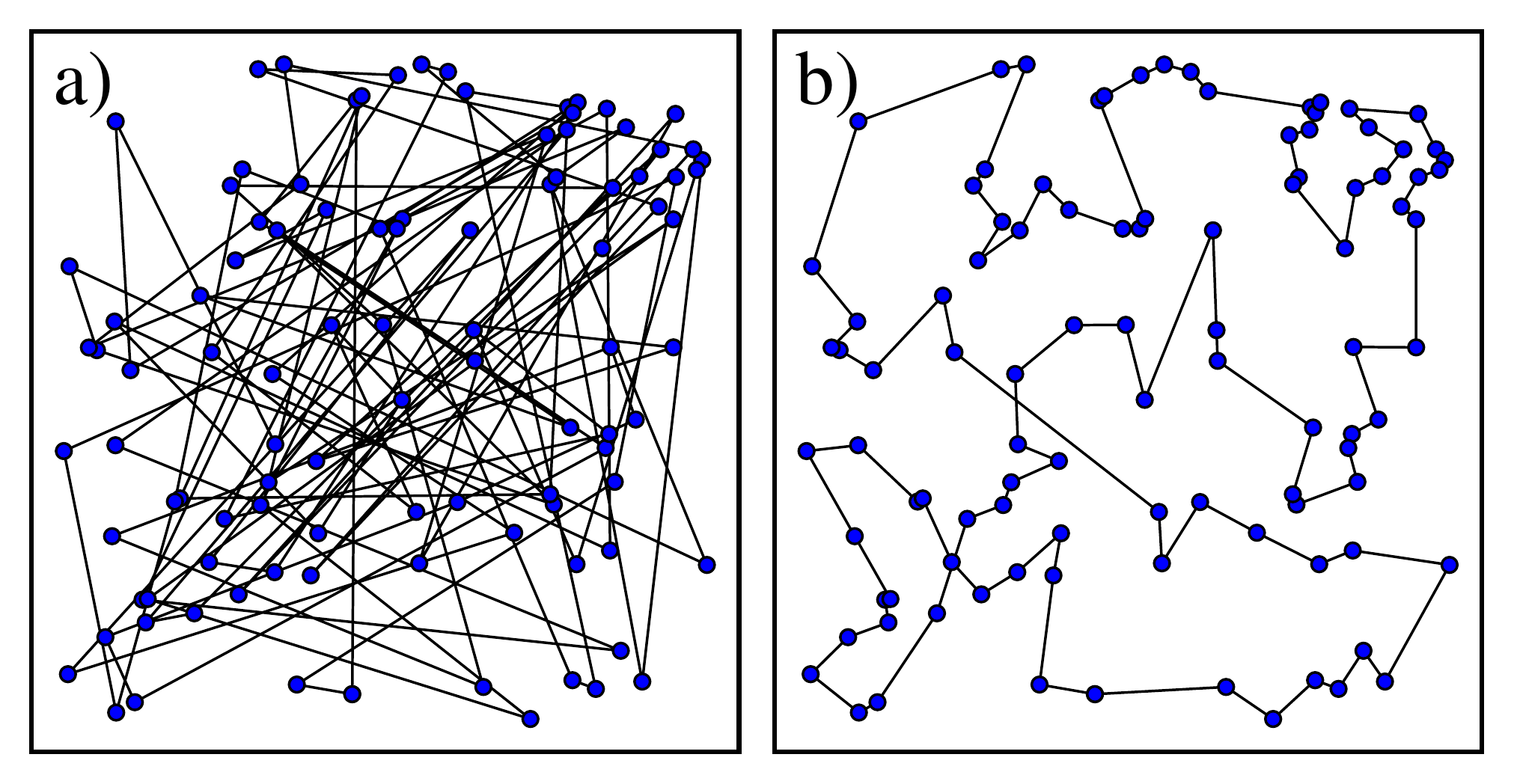}
	\caption{Sistema com 100 cidades distribuídas aleatoriamente. Figura~(a) mostra a configuração inicial do caixeiro viajante, rota com custo igual a $52,55$,  e (b) mostra o resultado do algoritmo, com custo reduzido para $8,56$  depois de $2\times10^6$ iterações. Parâmetros: $\alpha=0.99$, $\beta_{\textnormal{inicial}}=10$ e temperatura reduzida a cada $10^4$ passos.}
	\label{fig:100cities}
	\end{center}
\end{figure}

Já a figura~(\ref{fig:eList}) e a figura~(\ref{fig:100cities}) mostram os resultados do \textit{simulated annealing} para um problema do caixeiro viajante com 100 cidades distribuídas espacialmente pela distribuição uniforme de probabilidade. 
A figura~(\ref{fig:eList})  mostra  a energia das configurações em função do número de passos dados. A configuração é provável em um caso com temperatura muito alta, sendo assim no começo do processo as transições para estados de menor energia são as majoritariamente aceitas e, conforme o sistema se estabiliza, a flutuação da energia começa a se acentuar pois o sistema passa a transitar entre estados de energia próximas. 
Na figura~\ref{fig:eList}a), onde a diminuição da temperatura é maior, ou seja, a cada $10^4$ iterações ela diminuiu em $10\%$, notamos que ela atinge o valor mínimo da execução mais rapidamente, porém o sistema fica preso nesse estado metaestável, pois as flutuações não conseguem mais escalar a barreira de potencial. 
Por outro lado, a figura~\ref{fig:eList}b) mostra o caso onde a diminuição da temperatura foi de $1\%$ a cada $10^4$ passos, o que levou a um estado de menor energia. Podemos notar que ainda há flutuação e que uma energia menor ainda poderia ser atingida. 
A mesma distribuição de cidades foi usada nas figura~\ref{fig:circ}, porém com condições iniciais diferentes, o que não altera o resultado final da simulação. 

Notamos que a mudança no parâmetro da diminuição da temperatura, $\alpha$, altera a velocidade de convergência e a qualidade do  estado final.
Esse efeito, também é observado na mudança do número de passos dados antes da redução da temperatura, $\eta$, pois estão relacionados com a termalização do sistema e a adiabaticidade do processo como um todo. 
A qualidade do resultado da meta-heurística, em geral, vai depender desses dois parâmetros, que, por suas vezes, serão dependentes do tempo de relaxação\footnote{Tempo que o sistema demora para retornar ao equilíbrio termodinâmico} do sistema termodinâmico.  
Dessa maneira, uma possível otimização em algoritmos de \textit{simulated annealing} é a implementação de parâmetros dinâmicos, assim a temperatura é decrescida mais rápida quando a termalização é mais rápida, normalmente em temperaturas altas, e para tempos de relaxação mais lentos a temperatura é reduzida proporcionalmente a ele~\cite{de2003placement}. 

Mais adiante, o resultado da figura~(\ref{fig:100cities}) mostra o desempenho do SA que provavelmente não encontrou a melhor solução, mas foi preciso apenas 2 milhões de iterações. 
Como cada iteração do algoritmo tem complexidade da ordem de $\mathcal{O}(N)$, com $N=\text{'número de cidades'}=100$, foi preciso da ordem de $10^{8}$ operações. 
Como o algoritmo é heurístico seus parâmetros variam de problema para problema e o número de iterações depende da precisão em estudo, de forma que a complexidade computacional não pode ser determinada formalmente.  
Relembramos que para obter a solução ótima, o melhor algoritmo tem complexidade de $\mathcal{O}(n^2 2^n)$~\cite{held1962dynamic} e, assim teríamos que realizar da ordem de $100^2 2^{100} \approx 10^{33}$ operações, o que, mesmo com a propriedade de ser paralelizável, essa estratégia ainda seria muito mais custosa, apesar de ser exata. 

\section{Conclusão}

Neste trabalho, de fins pedagógicos, mostramos como a busca da solução de um problema de otimização pode ser representada como um processo markoviano, que é visto como um processo termodinâmico, e como isso junto à terceira lei da Termodinâmica são explorados pela meta-heurística \textit{simulated annealing}.
Mais especificamente, vimos a transformação do problema de otimização em um problema termodinâmico, e como o seu resfriamento até uma temperatura próxima de zero, onde em cada diminuição da temperatura o sistema é evoluído para o equilíbrio, se traduz na solução, geralmente aproximada desse problema.
Também foi discutido alguns problemas e características dessa técnica e o porquê dela não reproduzir a solução ótima em um caso genérico. 

Consideramos a construção de um algoritmo \textit{annealing} aplicado ao problema do caixeiro viajante e vimos que os resultados aproximados são satisfatórios no sentido de melhorar consideravelmente uma solução aleatória ou até encontrar a melhor solução para o problema. 
Também vimos como aplicar a técnica \textit{annealing} ao problema de Ising de forma que todos os problemas de otimização que recaem em um problema de Ising podem ser abordados de maneira análoga. 
Em síntese, como problemas NP, em geral, não podem ser resolvidos de forma exata por falta de recursos computacionais, os algoritmos \textit{simulated annealing} se mostram uma boa estratégia para encontrar boas soluções aproximadas rapidamente e com uma implementação simples.

Existem diversos estudos acerca da meta-heurística \textit{simulated annealing} que trazem discussões mais aprofundadas da teoria do método e apontam técnicas para melhoria de desempenho como a inclusão de parâmetros dinâmicos, outras maneiras de transitar entre estados e discussões acerca da condição de parada.
Também é possível desenvolver técnicas de paralelização para aumentar o seu desempenho~\cite{ram1996parallel}.
Outras discussões e técnicas sobre os assuntos tratados aqui podem ser encontradas nas referências~\cite{2019simulated, 1993simulated, penna1995traveling, tsallis1996generalized}.

\section*{Agradecimentos}
O autor agradece ao professor Celso Jorge Villas-Boas
pela orientação e pelas ótimas discussões. O autor agradece à Coordenação de Aperfeiçoamento de Pessoal de Níıvel Superior (CAPES/STINT), bolsa 88887.486234/2020-00, e ao Departamento de Física da UFSCar.


\end{document}